\documentclass[fleqn,twoside]{article}
\usepackage{espcrc2}

\usepackage{graphicx}
\usepackage[figuresright]{rotating}

\newcommand{\AmS}{{\protect\the\textfont2
  A\kern-.1667em\lower.5ex\hbox{M}\kern-.125emS}}

\title{A Possible Detection of Solar Gamma-Rays by the Ground Level Detector}

\author{Y. Muraki\address[STEL]{ Solar-Terrestrial Environment Laboratory, Nagoya University, 
        Nagoya 464-8601, Japan}%
        \thanks{email: muraki@stelab.nagoya-u.ac.jp},
  J. F. Valdes-Galicia\address[UNAM]{Instituto de Geofisica, Universidad Nacional
                 Autonoma de Mexico,04510, D.F. Mexico, Mexico\\}
  L. X. Gonzalez\addressmark,
  K. Koga\address[JAXA]{Aerospace Research and Development Directorate, JAXA, 
         Tsukuba 305-8505, Japan\\}
  H. Matsumoto\addressmark,
  S. Masuda\addressmark[STEL],
  Y. Matsubara\addressmark[STEL],
  Y. Nagai\addressmark[STEL],
  Y. Tanaka\address{Department of Physics, Hiroshima University, 
           Higashi-Hiroshima 739-0046, Japan \\},
  T. Sakai\address{College of Industrial Technologies, Nihon University, 
           Narashino 275-0005, Japan \\},
  T. Sako\addressmark[STEL],
  S. Shibata\address{Engineering Science Laboratory, College of Engineering, Chubu University,
            Kasugai 487-0027, Japan\\},
  and
  K. Watanabe\addressmark[JAXA]}
       
\begin{document}

\begin{abstract}
On March 7, 2011 from 19:48:00 to 20:03:00 UT, the solar neutron telescope 
located at Mt. Sierra Negra, Mexico (4,600 m) observed a 8.8{$\sigma$} enhancement.   
In this paper, we would like to try to explain this enhancement by a hypothesis 
that a few GeV gamma-rays arrived at the top of the mountain produced by the Sun.  
We postulate that protons were accelerated at the shock front.  
They precipitate at the solar surface and produced those gamma-rays.  
If hypothesis is confirmed, this enhancement is the first sample 
of GeV gamma-rays observed by a ground level detector. 
\vspace{1pc}
\end{abstract}

\maketitle

\section{Introduction}

 This paper describes an excess of the counting rate observed by the Solar Neutron Telescope (SONTEL) located at Mt. Sierra Negra in Mexico (4,600m a.s.l. or 575g/$cm^{2}$) 
on March 7, 2011 from 19:48 to 20:03 UT.  We attempt to explain this excess as due to the M3.7 solar flare observed by GOES.  The solar flare started at 19:43 UT and reached its maximum 
at 20:12 UT.   The RHESSI satellite also observed this flare from 19:27 UT to 20:08 UT.  
However the FERMI-LAT satellite and the SEDA-AP on board 
the International Space Station (ISS) were in the night side of the Earth until 20:02 UT.  

   According to the SEDA observation, the FIB neutron detector observed an excess induced by solar neutrons between 20:02 and 20:10 UT \cite{bib:muraki}.  On the other hand the FERMI-LAT satellite observed a long lasting GeV gamma-ray emission after 20:08 UT \cite{bib:tanaka}.  
The emission continued for more than 12 hours.  
The Solar Dynamical Observatory observed a Coronal Mass Ejection (CME) associated with this flare using their UV detector \cite{bib:cheng}.   
Therefore this is a good opportunity to propose a solution 
for the long standing question as to whether the flare starts first 
or the CME starts first and flares are produced afterwards \cite{bib:gosling}, and question is 
whether the long lasting emission of gamma-rays was induced by the continuous 
acceleration of the ions or by the precipitation of the trapped ions in 
the upper magnetic loop \cite{bib:ramaty}.  

  This paper is organized as follows: In Section 2, the solar neutron telescope 
located at Mt. Sierra Negra is described.  In Section 3, we provide the data 
obtained by five different kinds of satellites; the flare was observed by GOES, SEDA-FIB, FERMI-LAT, SDO and RHESSI satellites.   
A unified model is proposed to explain these observations in Section 4.  
Finally, in Section 5 we summarize and conclude.

\section{Solar Neutron Telescope at Sierra Negra}

A Solar Neutron Telescope (SONTEL) was installed at Sierra Negra on March 12, 2003.  
At Mt. Sierra Negra, Mexican astronomers were making a Large Milli-meter
radio Telescope (LMT) there.   
They installed a power line from a village in the foot of mountain. 
The authorities of INAOE 
(The Instituto Nacional de Astrofisica, Optica y Electronica) 
kindly accepted our proposal to use the power line for 
a cosmic ray detector and the construction of the detector started. 

\begin{figure}[t]
  \centering
  \includegraphics[width=0.50\textwidth]{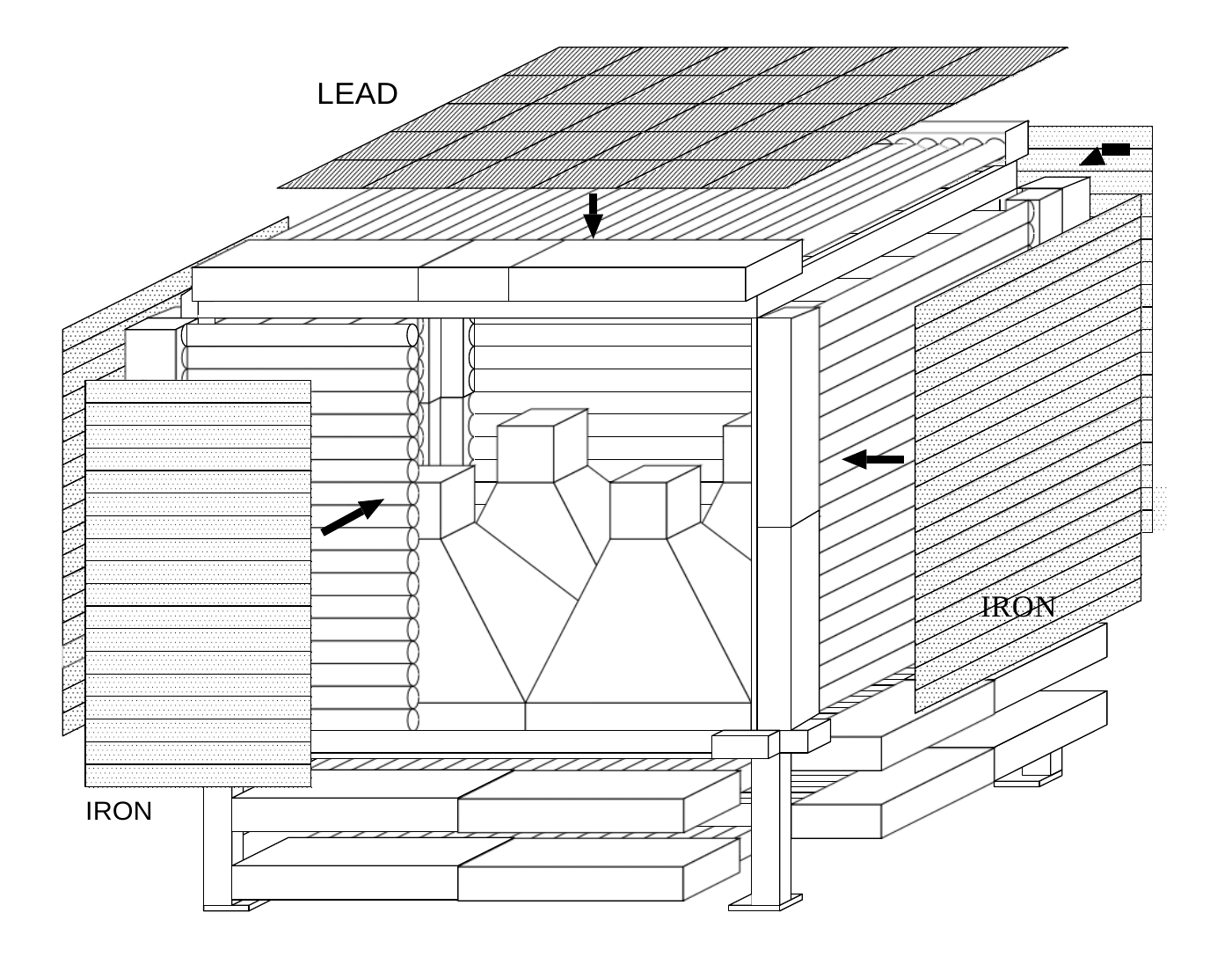}
  \caption{Solar Neutron Telescope at Mt. Sierra Negra. 
   The core of the detector consists of 30cm thick plastic scintillator 
   with an area of 4$m^{2}$.  The scintillator is surrounded by anti-counters. }
  \label{simp_fig}
 \end{figure}

As shown in Figure 1, the core of the detector consists of 
30cm thick plastic scintillator with an area of 4$m^{2}$.   
The scintillators are used for the measurement of the energy of neutron-induced-protons.   
The arrays of the proportional counters beneath the plastic scintillator 
are used for the determination of the direction of the incoming neutrons.   
Surrounding these detectors, proportional (anti) counters are located 
to identify the charged particles (e,  p, $\mu$) and 
separate them from the neutral particles (n, $\gamma$).  
At the top of the anti-counter, a 5 mm thick lead plate is installed to stop 
the flux of gamma-rays.  However, 
the efficiency of the anti-counter is estimated to be about 80{$\%$} 
due to lack of coverage (gap) and the conversion rate 
of photons into electron-positron pairs in the lead and iron plates.  

Here we show typical trigger rate for 10 years in Table I, separating them 
into three periods; 2003, 2005, and 2011.   The trigger rates of anti-counters (anti-all) 
and scintillators have been quite stable for 10 years, however slight increase 
of the counting rate for neutral channels S1 and L1 is recognized.  
Here S1$\_$charged denotes incident charged particles (e, p, $\mu$) with deposited energy 
in the scintillator higher than 30 MeV and S1 represents the counting rate with 
the same deposited energy but induced by neutral particles (n, $\gamma$).   
L1 represents the counting rate of the top array of the PR counters 
underneath the plastic scintillator.   

The counting rate of the L1 channel should be produced by penetrating charged particles 
from the scintillator like protons or electrons, 
originated from neutral particles like neutrons or gamma-rays (n, $\gamma$).  
The S1 (neutral) represents the counting rate induced 
by neutral particles inside the scintillator.    
At the moment, we do not understand the reason of the increase;
probably it arises from the degradation of one of the IC chips used to make the trigger 
logic; Xlink Complex Programmable Logic Device (CPLD), due to the radiation damage 
at high altitude \cite{bib:hirano}.  Another possibility may arise from the increase 
of the flux of gamma-rays during solar cycle 24.  Anyway the ratio S1/L1 increased only 9$\%$
from 2005 to 2011.

\begin{table}[h]
\begin{center}
\begin{tabular}{|l|c|c|c|}
\hline 
       channel & 2003.11 & 2005.9  & 2011.3    \\ \hline
      anti-all  &  687,000   &  707,000 & 749,000  \\ \hline
     S1-charged  & 186,000   & 171,900  & 173,400   \\ \hline
     S1(neutral)& 67,800   & 67,000  & 85,000    \\ \hline
     L1(neutral)& ---      & 23,300  & 27,030   \\ \hline
\end{tabular} 
\caption{Typical counting rate of respective year (one minute value).  
         From top to bottom, the counting rate of anti-all, S1(charged),
         S1(neutral) and Layer 1(L1:neutral) respectively.  }
\label{table_single}
\end{center}
\end{table}

\section{The time profile of the S1 and L1}
n Figures 2(a) and 2(b), we present the time profile 
of the two typical neutral channels, S1 and L1.   
The excess of the counting rate at the L1 channel can be identified around 19:48-20:03 UT.  
It is very clear and the statistical significance of the excess was calculated 
as to be 8.8$\sigma$ using Li-Ma statistics.  The number of excess was 5,700 counts within 15 minutes.   On the other hand, the excess of S1 channel was about 6.0$\sigma$ and the number 
of counts was about 6,800 events.  
What is the problem?   According to our observation of solar neutrons by the same detector on September 7, 2005, the ratio between S1/L1 was 10 \cite{bib:sako}; however, 
in this case the ratio is only 1.2$\sim$1.8.   Can we explain this ratio reasonably?   
Before we enter the discussions on this matter, let us look at the 
space environment through the data collected by several satellites.

\begin{figure*}[!t]
\centering
\includegraphics[width=0.78\textwidth]{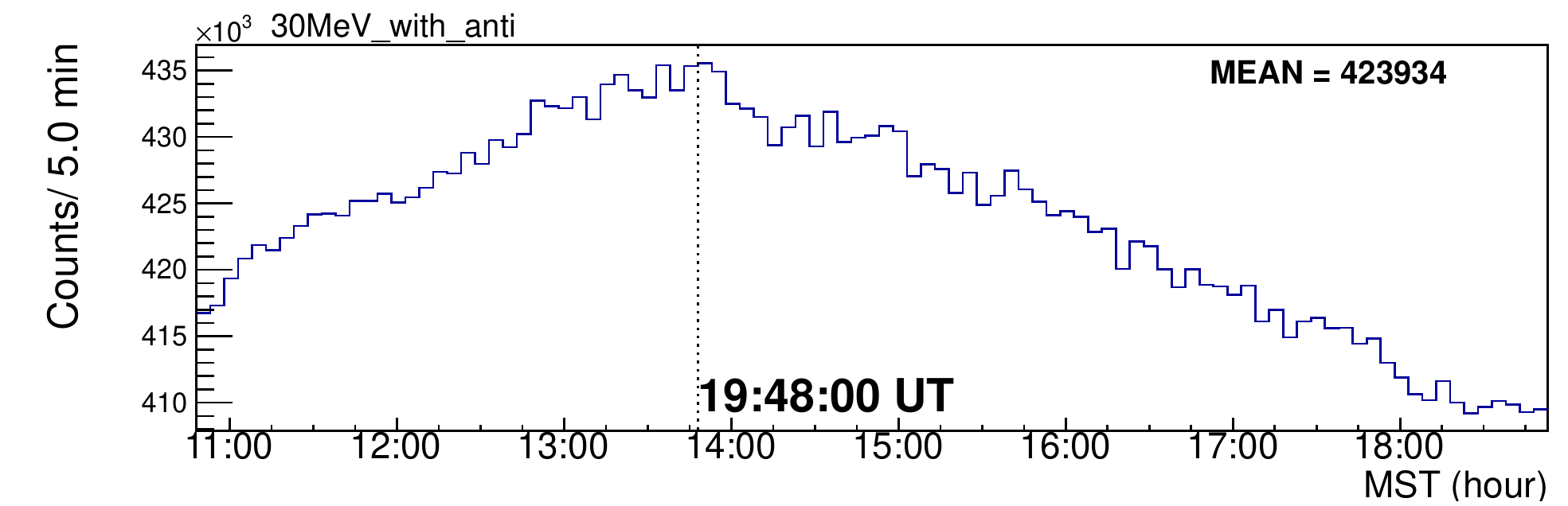}
 \caption{Counting rate of S1 (5 minutes value).  Neutron-induced-proton channel
  with the deposited energy higher than 30 MeV. }
 \label{simp_fig}
 \end{figure*}
 
\begin{figure*}[!t]
 \centering
 \includegraphics[width=0.78\textwidth]{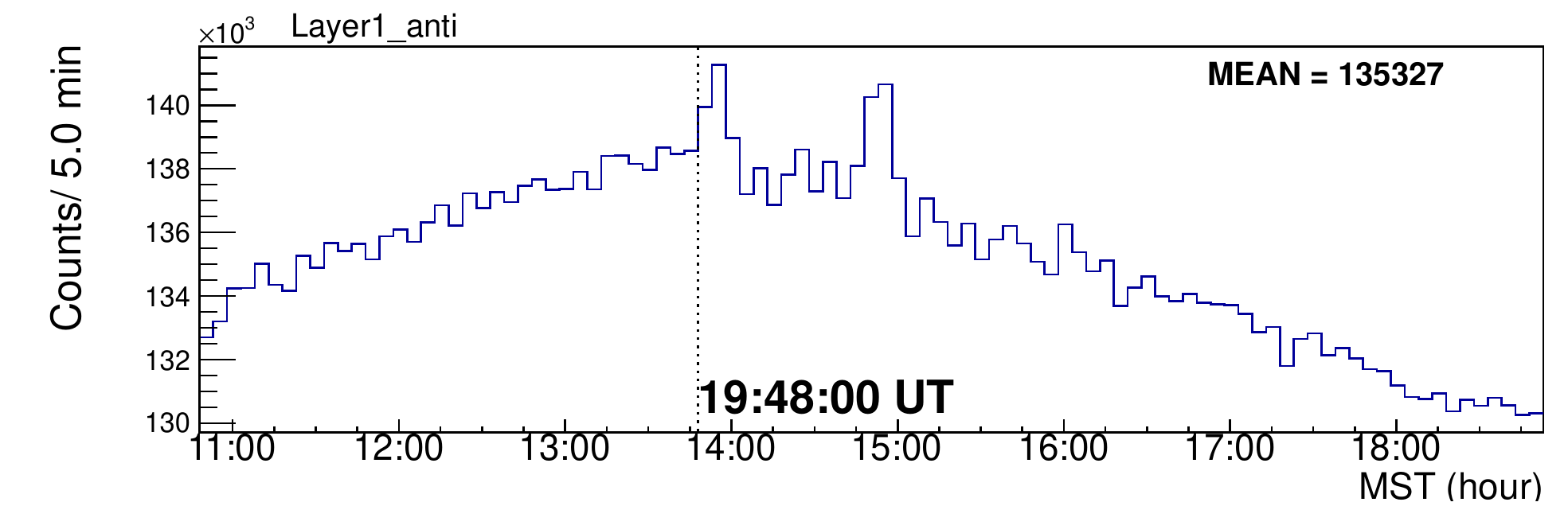}
 \caption{Counting rate of S1 and L1 (5 minute value).  Incident particles are
 neutral.  They deposit the energy higher than 30 MeV in the scintillator.  
 The L1 channel represents the charged particles penetrating the scintillator
  and detected by the top layer of the PR-counters.}
 \label{simp_fig}
\end{figure*}

\section {Space Environment Data  }

Here we will present the space environment data obtained around 20 UT on March 7, 2011, based on five different satellites; GOES, RHESSI, FERMI-LAT, SEDA-FIB and SDO satellites.

1. {$\it GOES$} data

The GOES satellite observed a flare of magnitude of M3.7.  The flare started at 19:43 UT and reached its maximum at 20:13 UT.  
The flare position was identified at about N23W50 in Region 1164. 

2. {$\it RHESSI$} satellite

The RESSHI satellite observed a solar flare from 19:27 UT to 20:08 UT.   The intensity at the beginning was weak, we can therefore make a two dimensional plot of the hard X-rays at   19:57 UT and 20:01 UT.  They are presented in Figures 3a and 3b.  We can recognize from these two pictures, that the position of the flare moved between the two times. 

\begin{figure}[t]
  \centering
  \includegraphics[width=0.45\textwidth]{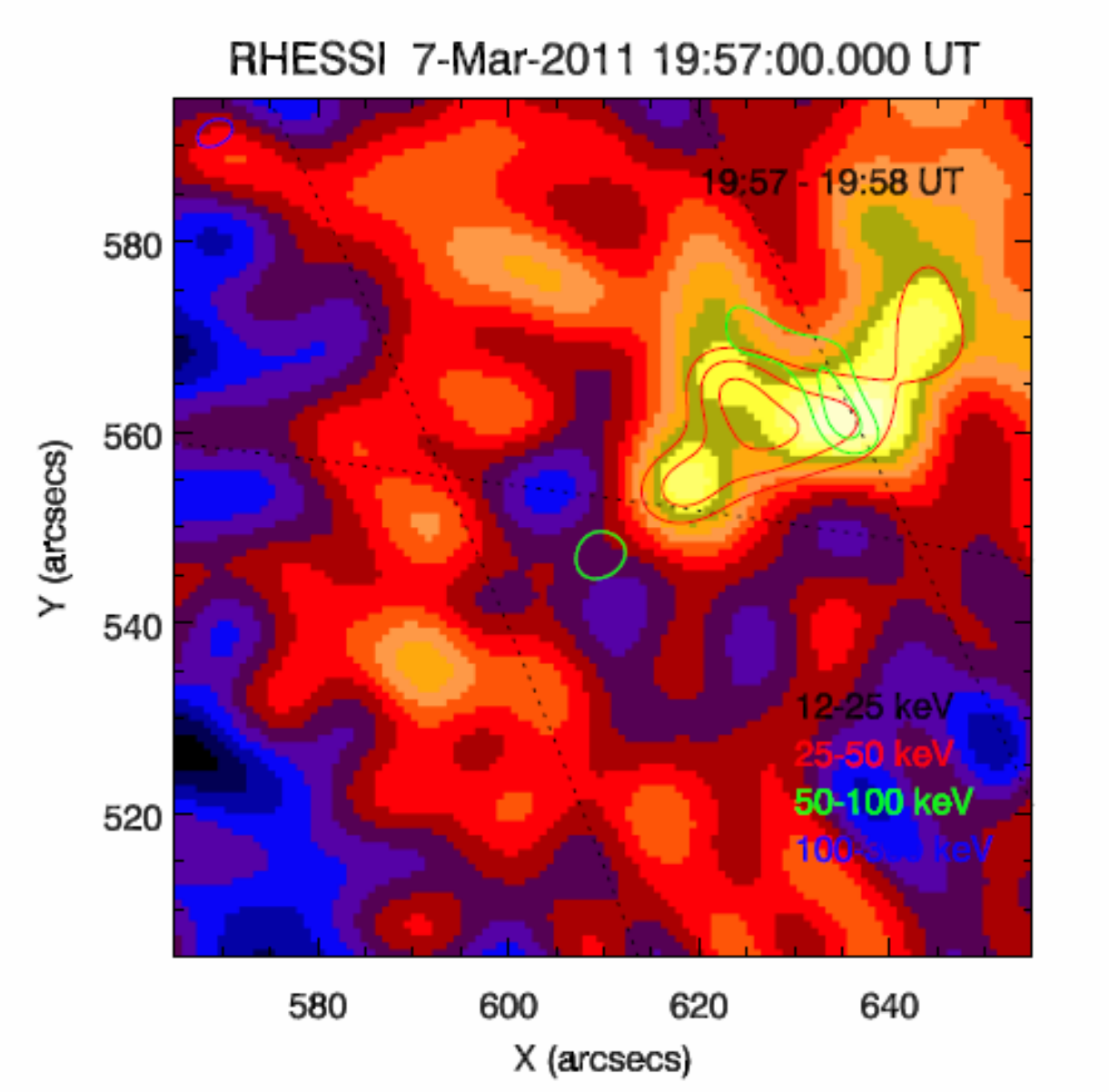}
 
  \includegraphics[width=0.45\textwidth]{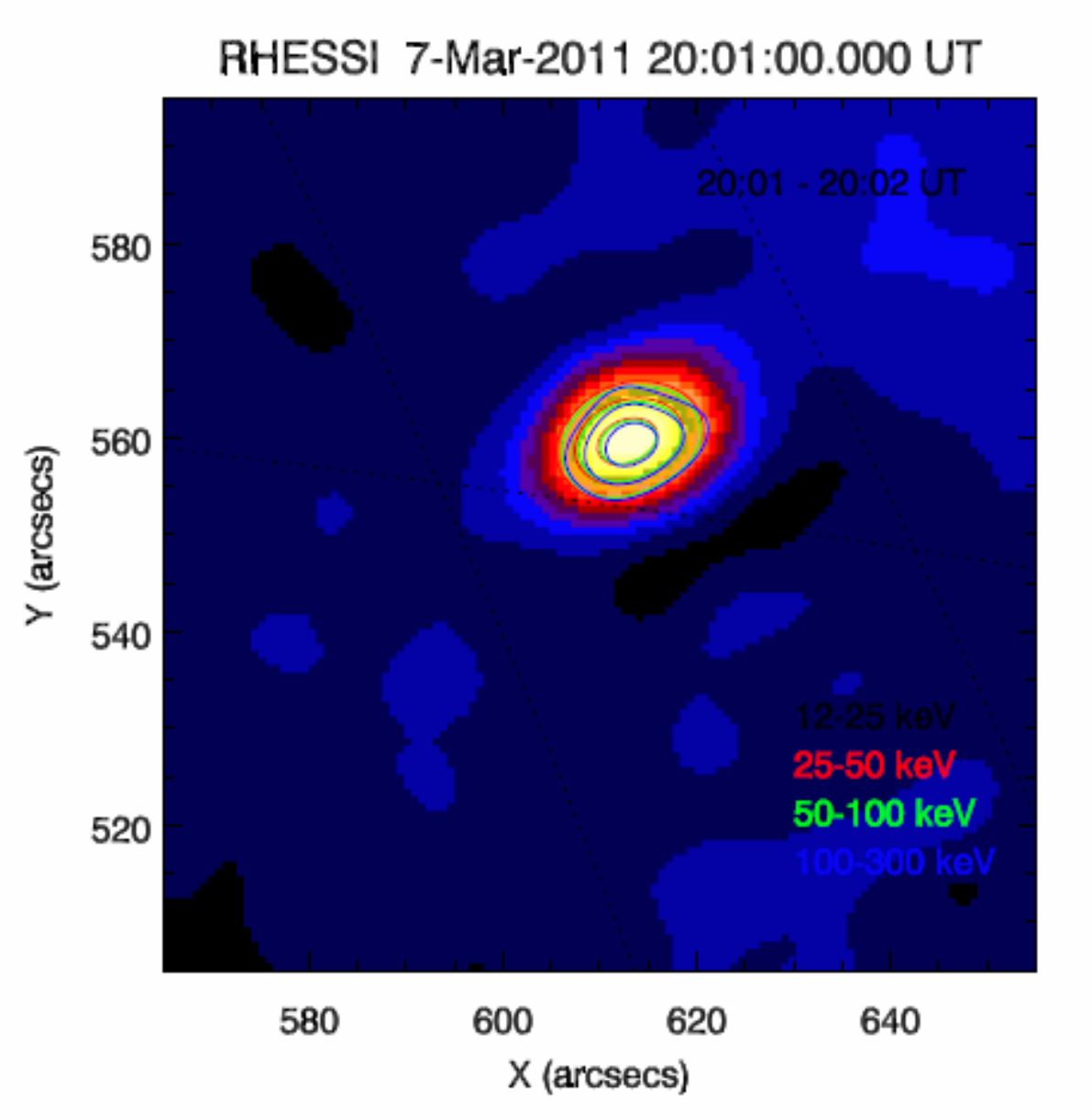}
  \caption{(top) RHESSI 19:57-19:58 UT data and
   (bottom) 20:01-20:02 UT data. The red contour represents the
    intensity of the hard X-rays with 25-50 keV, while
    the green contour corresponds to soft gamma-rays
    with the energy of 50-100 keV. }
  \label{simp_fig}
 \end{figure}

3. {$\it FERMI-LAT$} satellite

The FERMI-LAT satellites observed a flare from 20:08 UT. Gamma-rays with a peak intensity 
at 200 MeV was observed.  The spectrum extended beyond 1.5 GeV.  
The gamma-ray emission continued for 12 hours \cite{bib:tanaka}. 

4. {$\it SEDA-FIB$} detector on ISS

The SEDA-AP is a neutron sensor on board the International Space Station.  The sensor can measure the energy and arrival direction of neutrons.  Therefore the production time can be estimated through the information on the energy of neutron-induced-protons.   Figure 4 indicates the production time of those neutrons.  The most probable production time of neutrons is estimated as to be around 19:48 UT \cite{bib:muraki}.

\begin{figure}[t]
  \centering
  \includegraphics[width=0.45\textwidth]{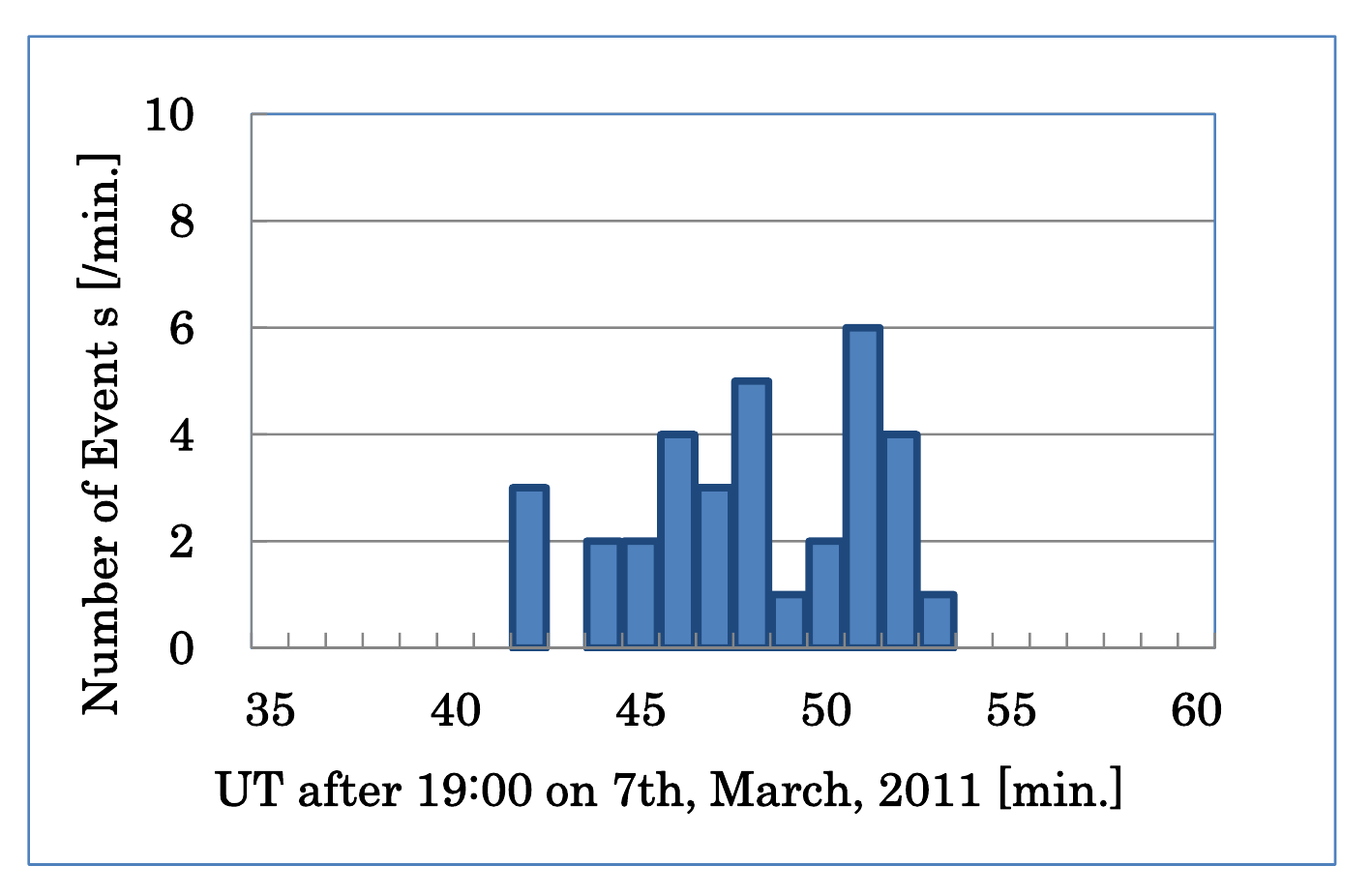}
  \caption{Departure time of solar neutrons at the Sun
    estimated by the SEDA-FIB neutron data.   They
    seem to have departed at $\sim$19:48 UT. }
  \label{simp_fig}
 \end{figure}

5. {$\it SDO$} satellite

The Solar Dynamical Observatory is looking at the Sun by means of ultra-violet telescopes over different wave lengths.  The SDO satellite succeeded to observe the emission of a Coronal Mass Ejection (CME) associated with this flare from the emission of early times.  
Furthermore not only they succeeded to observe the birth of the CME but also its development together with the start of an impulsive flare at 19:43 UT.   
In Figure 5a (extension of CME) and 5b (start of impulsive flare), 
we show the data described above.   
They are taken at 19:47 UT and 19:59 UT respectively.

 
 \begin{figure}[t]
  \centering
  \includegraphics[width=0.45\textwidth]{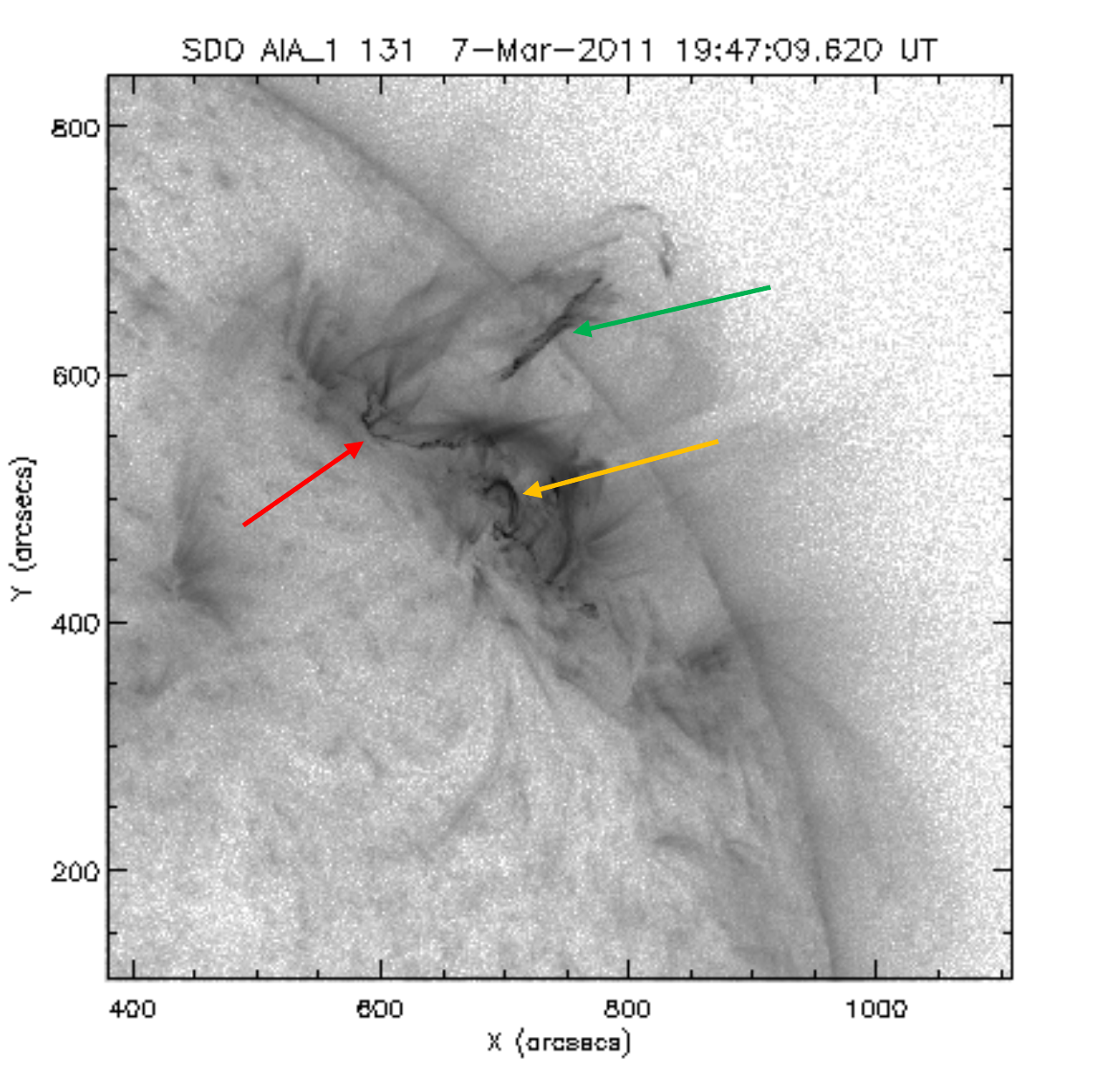}
 
  \includegraphics[width=0.45\textwidth]{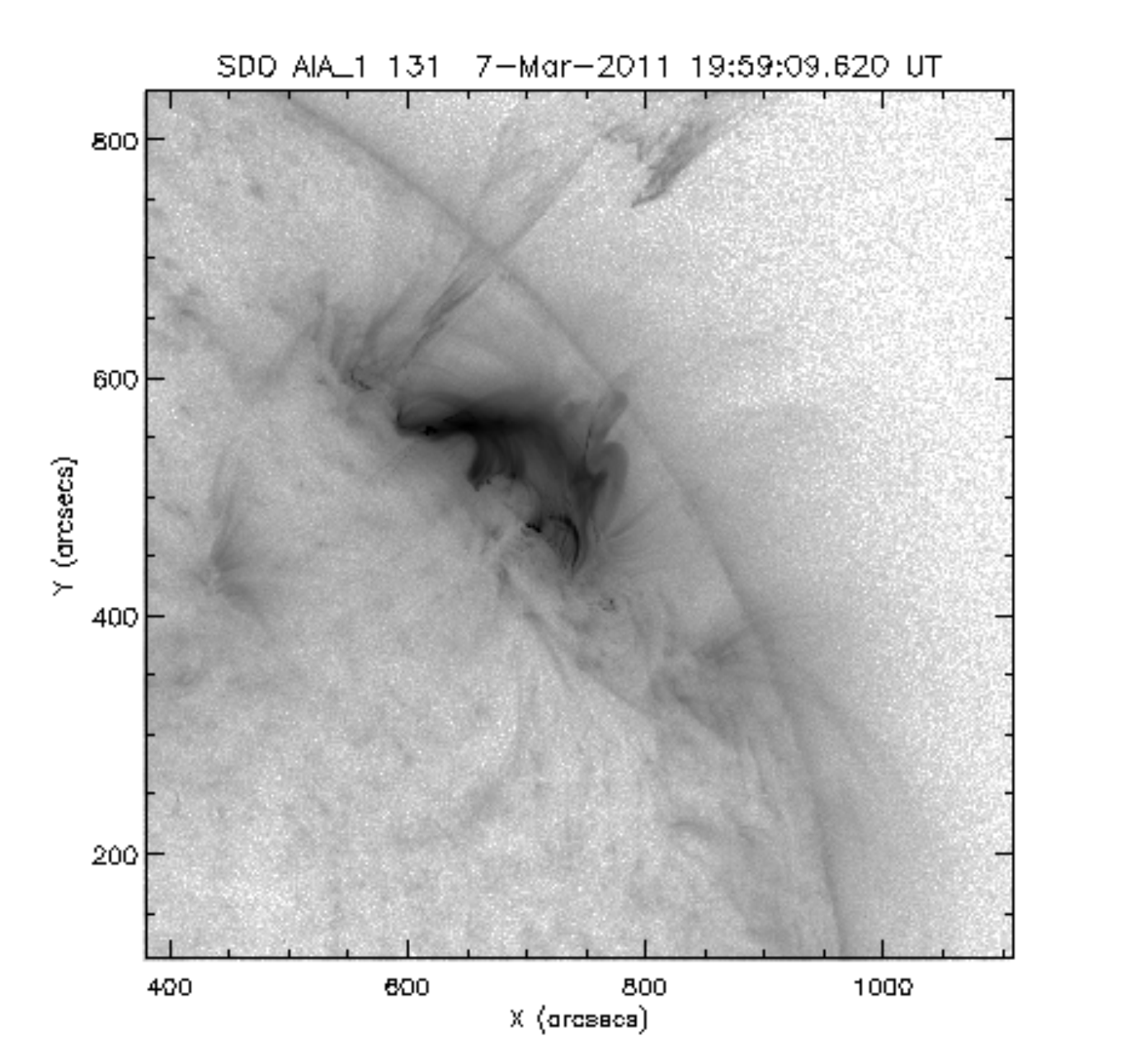}
  \caption{The flare and CME observed by the UV telescope of SDO (131 nm).
   The top view shows the expansion phase of the CME.  The bottom photo
   was taken at the time of the impulsive flare.  Arrows red, green and orange represent
   the birth place of the CME, the acceleration region of
    protons and impulsive flare region.  From top view
    a transportation of hot plasma into the foot point
    of the CME (indicated by the red arrow) can be recognized. }
  \label{simp_fig}
 \end{figure}

In the next, we will introduce a hypothesis and try to explain in search of a unified model  
for the whole event.


\section {Interpretation for the Sierra Negra event}
\subsection{A possible gamma-ray detection}
We provide here an interpretation of the excess on the counting rate observed 
at Mt. Sierra Negra.   First of all, the observed ratio of S1/L1 was 1.2$\sim$1.8, 
depending on how estimate the background.   However,  
the result sharply contrasts with the assumption that this excess was produced by neutrons, 
as for the observation of solar neutrons on September 7, 2005, this ratio was 10.  
These observations have been extensively discussed in Watanabe et al 2006 \cite{bib:watanabe1} 
and Sako et al 2007 \cite{bib:sako}.  
According to the calculations \cite{bib:watanabe2}, the ratio was predicted to be 8 at En=200 MeV 
and 20 at En=100 MeV.   However the same MC calculation predicts for gamma-rays a ratio 
of 1.5 at 200 MeV and 3 at 100 MeV.  

We have made a new full MC calculation
for gamma-rays with different energies.   As a result of
MC calculations, the value turns out to be $\sim$3 between E$\gamma$=1 GeV to 10 GeV.
Based on this argument, we believe that this excess was made by gamma-rays.
New MC calculation also shows the ratio of S1/L1 for neutron incidence to be $\sim10$.  

The radiation length of the scintillator is about 40 g/{$cm^{2}$}.  Therefore if gamma-rays enter
into the plastic scintillator, they will be converted into electron-positron pairs with
high probability.  However those electron and positron pairs are passing through the
scintillator with the minimum deposited energy like muons.   On the contrary, protons
loose the energy in the scintillator.   This makes a sharp contrast between proton passage
and electron passage in the scintillator. Thus they produce less trigger rate 
of the L1 channel for protons.
  
Can gamma-rays arrive at mountain detectors?  
When high energy gamma-rays enter into the atmosphere, electro-magnetic showers are produced.  
The incident energy is distributed into individual energies of secondary photons 
and electron-positron pairs.  
Therefore it is expected to be absorbed in the atmosphere before arrival of the detector.  
The critical energy to penetrate into the atmosphere is 84.4 MeV and the radiation length 
is 37.7g/{$cm^{2}$}.  
When very high energy gamma-rays enter into the atmosphere, 
they produce large air showers.   Secondary electrons and gamma-rays arrive 
at the top of the detector.   
We made a MC calculation for gamma-rays with energy range
from 50 MeV to 20 GeV.    
According to our MC calculation, when the incident energy of gamma-rays exceeds 3 GeV, 
the detection efficiency of gamma-rays by SONTEL exceeds beyond 1.0 (100$\%$).  
However for gamma-rays with E{$\gamma$}=1 GeV, the value is expected to be 0.2 (2$\%$). 
Concerning the S1/L1 ratio, the value is expected to be $\sim$3 
in a wide energy range of E$\gamma$=1 GeV - 20 GeV. 

Taking into account these circumstances, it would be natural 
to set the assumption that these excesses were produced by 
arriving high energy gamma-rays produced at the solar surface.   
If this is true, as far as we know, this is the first evidence 
that solar gamma-rays are detected by a ground level detector.  
Until now the detection of high energy gamma-rays have been made 
only at space vehicles.

\subsection{Production time and Mechanism}
  According to RHESSI, the intensity of hard X-rays (25-50 keV) reached its maximum at 20:02 UT.  The time does not correspond to the excess time of SONTEL (19:48-20:03 UT).   
On the other hand, solar neutrons were observed by SEDA-FIB.  
Solar neutrons were estimated to start from the Sun around 19:48 UT (Figure 4).   
This time accords with the enhanced time of SONTEL counting rate.  

We have carefully investigated the movies of the solar surface obtained by SDO.   
Some key pictures are given in Figures 5a and 5b.  
At 19:48 UT, CME already arched.  
Therefore we reached the conclusion that around 19:45 the 
hot plasma was injected into the magnetic loop of the CME.   
Those {$\it seed-protons$} were accelerated into high energy at the shock front 
of the CME by the shock acceleration mechanism.  
The highest energy of protons already reached over the energy of 30 GeV at 19:48 UT.    
When they reached high energy, the magnetic loop of 
the CME cannot involve those high energy protons inside the magnetic loop.  
So they precipitate onto the solar surface and produced a few GeV gamma-rays and neutrons.
The whole scenario may be summarized in Figures 6a and 6b.
The M3.7 solar flare was made at different place near the CME.

\begin{figure}[t]
  \centering
  \includegraphics[width=0.48\textwidth]{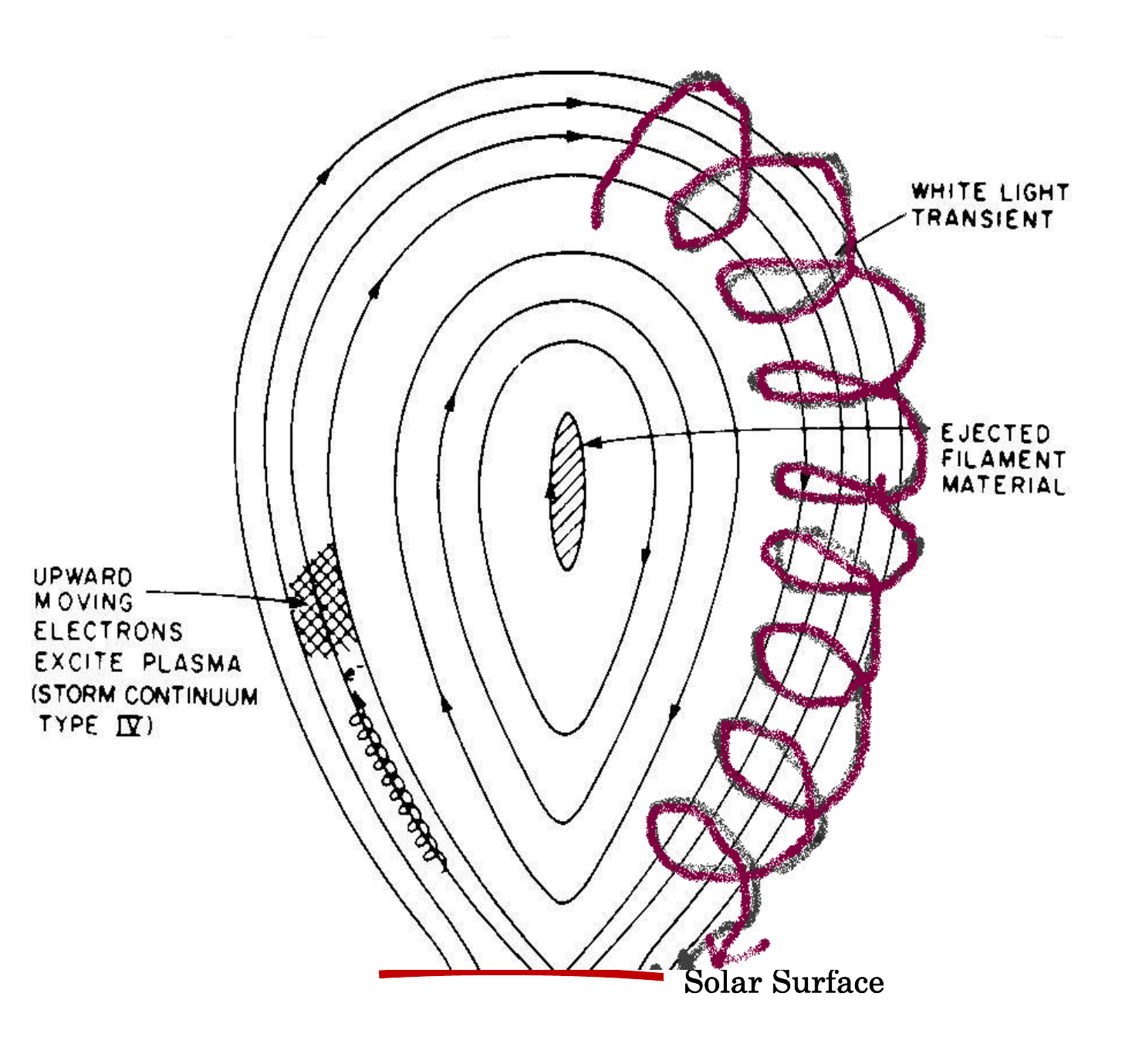}
  
  \includegraphics[width=0.30\textwidth]{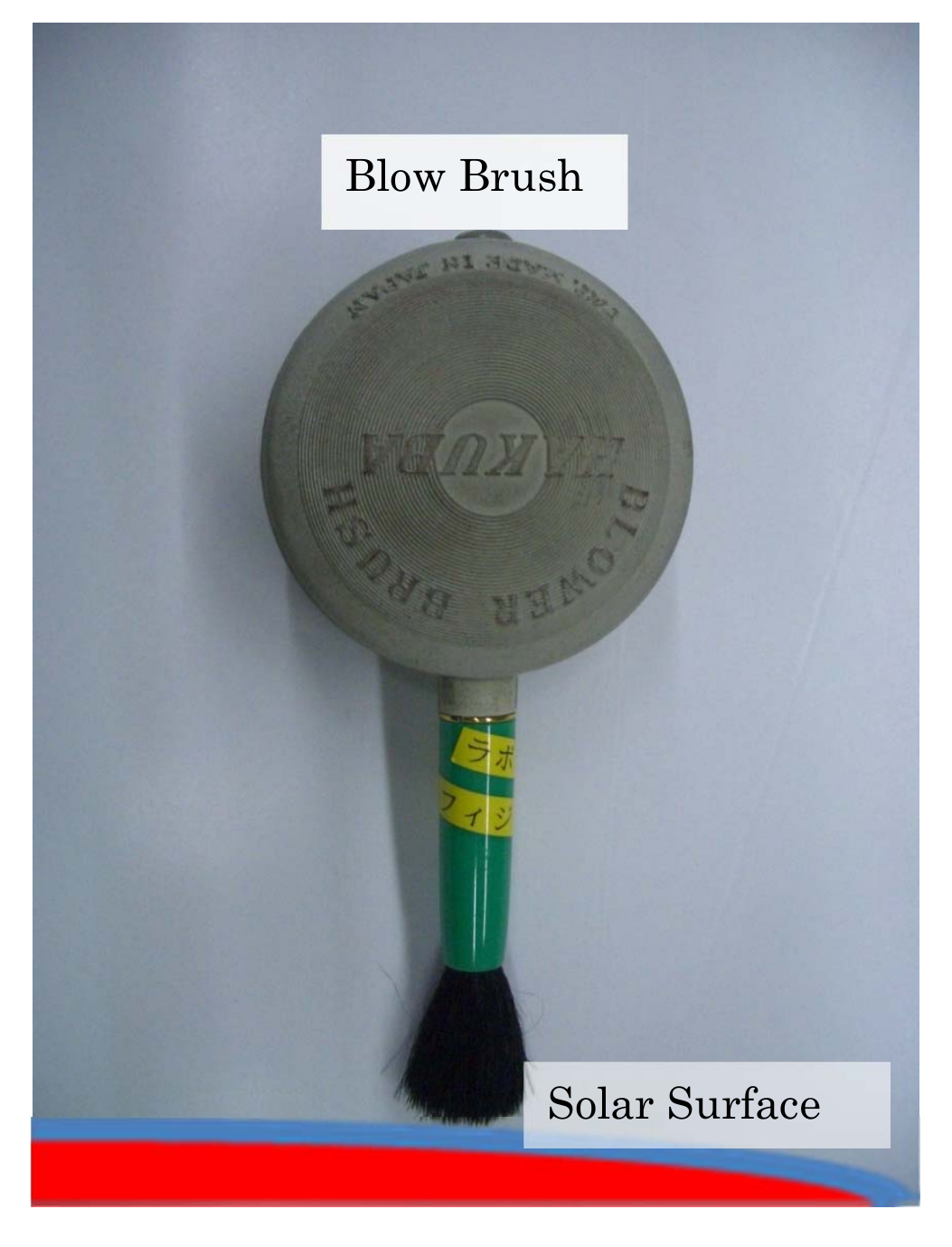}
  \caption{(a)top: Production scenario of high energy solar gamma-rays
   in connection with acceleration point in CME and production place
    of gamma-rays. The basic drawing is by the courtesy of E.W. Cliver
    (from AIP conference proceeding 374 (1995) page 53.)
   (b)bottom: A blow brush model for neutron and gamma production.
   Accelerated particles in the shock front are stored in the
   magnetic bubble and are precipitating over the solar surface.
   Then by nuclear interactions with the solar atmosphere, pi-zero and neutrons are
   produced. }
  \label{simp_fig}
 \end{figure}

\section{Conclusions}

 The excess observed by the SONTEL at Mt. Sierra Negra must be produced by 
high energy solar gamma-rays.   As far as we know, this is the first experimental data 
that solar gamma-rays detected by ground level detector.


\end{document}